\documentclass[prl,superscriptaddress,showpacs,longbibliography,reprint]{revtex4-1}
\usepackage[utf8]{inputenc}
\usepackage{amsmath}
\usepackage{amsfonts}
\usepackage{amssymb}
\usepackage{braket}
\usepackage{dsfont}

\newcommand{\G}{\mathcal{G}}

\newcommand{\ketbra}[2]{\ket{#1}\hspace{-2.1pt}\bra{#2}}

\DeclareMathOperator{\Tr}{Tr}

\newcommand{\nb}{\overline{n}}

\usepackage{graphicx}

\usepackage{color,soul}

\begin{document}
\title{Using $(1+1)D$ Quantum Cellular Automata for Exploring Collective Effects in Large Scale Quantum Neural Networks}

\author{Edward Gillman}
\affiliation{School of Physics and Astronomy, University of Nottingham, Nottingham, NG7 2RD, UK}
\affiliation{Centre for the Mathematics and Theoretical Physics of Quantum Non-Equilibrium Systems,
University of Nottingham, Nottingham, NG7 2RD, UK}

\author{Federico Carollo}
\affiliation{Institut f\"{u}r Theoretische Physik, Universit\"{a}t T\"{u}bingen, Auf der Morgenstelle 14, 72076 T\"{u}bingen, Germany}

\author{Igor Lesanovsky}
\affiliation{School of Physics and Astronomy, University of Nottingham, Nottingham, NG7 2RD, UK}
\affiliation{Centre for the Mathematics and Theoretical Physics of Quantum Non-Equilibrium Systems,
University of Nottingham, Nottingham, NG7 2RD, UK}
\affiliation{Institut f\"{u}r Theoretische Physik, Universit\"{a}t T\"{u}bingen, Auf der Morgenstelle 14, 72076 T\"{u}bingen, Germany}

\begin{abstract}
Central to the field of quantum machine learning is the design of quantum perceptrons and neural network architectures. 
A key question in this regard is the impact of quantum effects on the way in which such models process information. 
Here, we approach this question by establishing a connection between $(1+1)D$ quantum cellular automata, which implement a discrete nonequilibrium quantum many-body dynamics through the successive application of local quantum gates, and recurrent quantum neural networks, which process information by feeding it through perceptrons interconnecting adjacent layers. This relation allows the processing of information in quantum neural networks to be studied in terms of the properties of their equivalent cellular automaton dynamics. We exploit this by constructing a class of quantum gates (perceptrons) that allow for the introduction of quantum effects, such as those associated with a coherent Hamiltonian evolution, and establish a rigorous link to 
continuous-time Lindblad dynamics. We further analyse the universal properties of a specific quantum cellular automaton, and identify a change of critical behavior when quantum effects are varied, demonstrating that they can indeed affect the collective dynamical behavior underlying the processing of information in large-scale  neural networks.
\end{abstract}

\maketitle

\begin{figure}[t]
\centering
\includegraphics[width=1\linewidth]{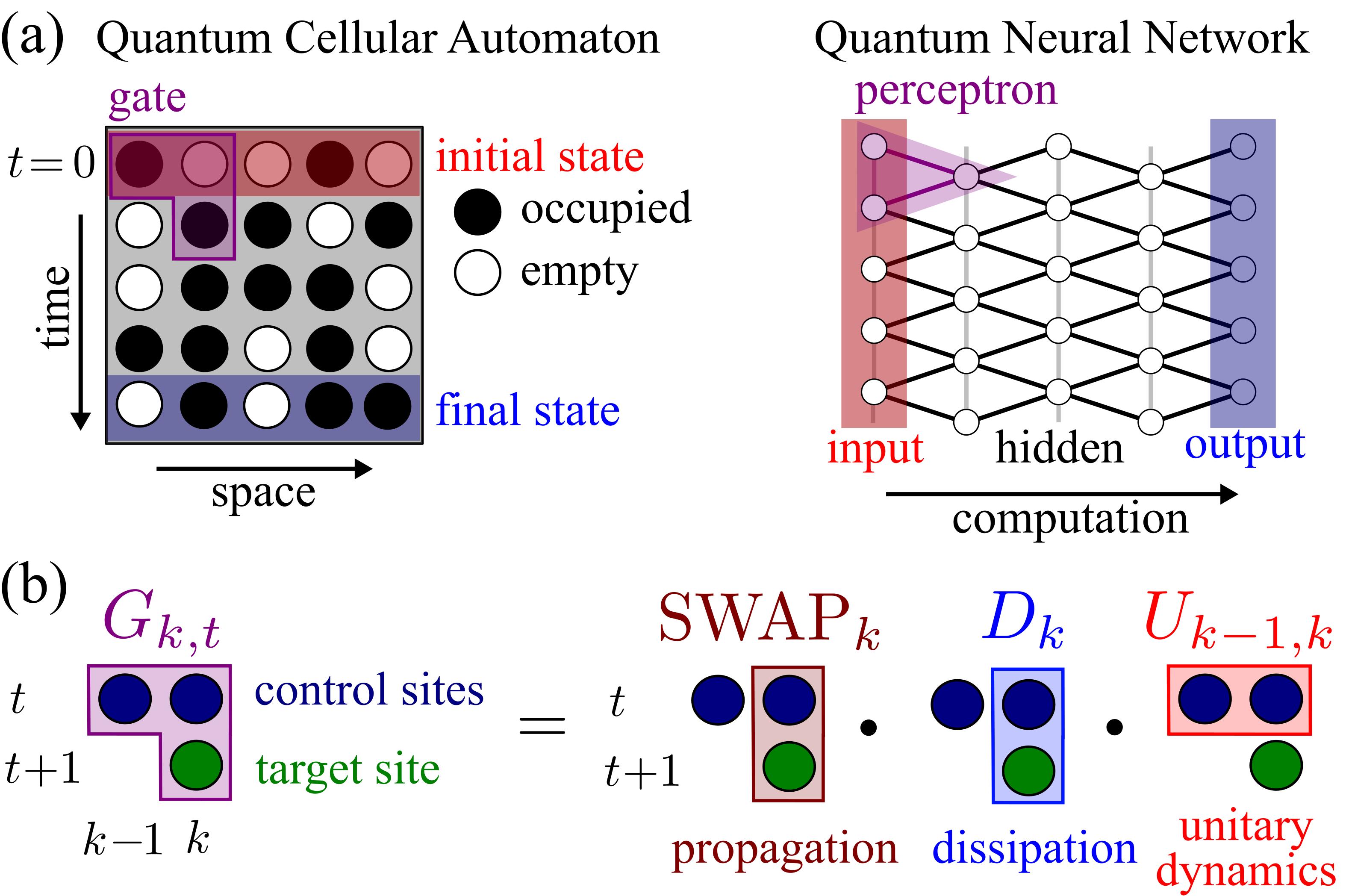}
\caption{\textbf{QCA and QNNs} \textbf{(a)} $(1+1)D$ QCA propagate an initial quantum state, defined on a one-dimensional lattice, along a discrete time direction via the application of quantum gates. First, one applies the gates successively to all qubits (sites) of a given row. This completes one time step and one subsequently advances to the next row. This discrete time evolution is closely related to the dynamics of a computation on a recurrent QNN. Here information is processed by perceptrons that link adjacent layers, thereby propagating the input (initial state) to the output (final state). \textbf{(b)} We consider a gate $G_{k,t}$ with two control qubits (on time slice $t$) and one target qubit (on time slice $t+1$). For the shown decomposition of the gate one obtains an effective open system dynamics (with single-site dissipation) through unitary gates: first, a unitary $U_{k-1,k}$ is applied to control qubits, which implements a Hamiltonian evolution. Second, an entangling gate $D_k$ acts on one of the controls and on the target qubit, resulting in dissipation. Finally, a propagation step which is implemented by applying a swapping operation, $\text{SWAP}_k$, between control and target qubits.}
\label{fig:QCA_models}
\end{figure}

An established concept for classifying phases of matter and transitions among them is that of universality. This builds on the observation that, near a critical point, different systems display macroscopic behavior that does not depend on their microscopic details \cite{goldenfeld2018}. Consequently, a wide variety of systems exhibit the same collective scaling behavior for key observables. In equilibrium this can be observed, e.g., in the magnetization of the Ising model and the specific volume of a van-der-Waals gas, both of which follow the same scaling laws near the critical point. In the realm of nonequilibrium phase transitions (NEPTs) universality can be, for instance, observed in dynamical processes that feature an absorbing state -- a particular configuration of the system which, once reached, causes the dynamics to halt. These systems feature emergent collective behavior that typically belongs to the so-called directed percolation (DP) universality class. This is, e.g., the case for both the continuous-time classical contact process (CCP) \cite{Harris1974, Henkel2008} and the discrete-time Domany-Kinzel cellular automata (DKCA) \cite{Henkel2008,Henkel2010,Domany1984, Bagnoli2001,Bagnoli2014,Hinrichsen2000,Lubeck2005}, which, despite their very different microscopic formulation, display the same quantitative large-scale critical physics. 

Recently, it has been found that quantum effects can lead to a change of the universal behavior of nonequilibrium stochastic processes. This has been observed in a continuous-time open quantum version of the contact process model \cite{Marcuzzi2016,Buchhold2017,Roscher2018,Carollo2019,Gillman2019,Jo2021}, with a similar change also found within so-called $(1+1)D$ quantum cellular automata (QCA) \cite{Gillman2022}, see Fig.~\ref{fig:QCA_models}(a). QCA represent an extension of classical cellular automata (CCA) into the quantum domain \cite{Lesanovsky2019, Gillman2020, Gillman2021, Gillman2021b, Nigmatullin2021} and can, for instance, be realized on quantum hardware based on Rydberg atoms \cite{Wintermantel2020, zeiher2016,kim2018,browaeys2020,ebadi2020}. Moreover, QCA are also closely related to quantum neural networks (QNNs) [see Fig.~\ref{fig:QCA_models}(a)], which are of considerable interest in current quantum machine learning research \cite{Beer2020,Patti2021,OrtizMarrero2021,Sharma2022}. This interest is driven by the idea  that QNNs might accomplish learning tasks related to quantum data more efficiently than their classical counterparts \cite{Schuld2022}. 

Building on the connection between $(1+1)D$ QCA and a paradigmatic class of QNNs \cite{Gillman2022,Beer2020,Sharma2022,Broughton2020}, we address here the question of how quantum effects impact information processing in {\it large} QNNs, which have the capability to develop emergent dynamical behavior. To this end, we first construct a class of quantum perceptrons (gates) such that the information processing of the QNN can be understood as a dissipative many-body quantum dynamics, that in some limit can be approximated by a continuous-time Markovian open quantum time evolution. We then investigate the impact of quantum effects on collective behavior of this dynamics by focusing on a setting where it displays an NEPT. Using numerical simulations based on tensor networks (TNs), we investigate long propagation times --- corresponding to deep QNNs --- and large system sizes --- corresponding to a large number of neurons per layer. We find that, indeed, quantum fluctuations can impact on emergent properties within QNNs, i.e., as a function of the parameters of the perceptrons they lead to changes in the universal critical dynamics.

\noindent \textbf{Relationship between QCA and QNNs.} A $(1+1)D$ QCA consists of a $2$D lattice of sites, each one having as basis states an {\it empty} one, $\ket{\circ}$, and an {\it occupied} one, $\ket{\bullet}$. 
Similarly to CCA, the whole lattice is prepared with all sites in state $\ket{\circ}$, apart from those in the first row --- here denoted as row $t=0$ as shown in Fig.~\ref{fig:QCA_models}(a). The latter encode the initial state of an effective one-dimensional system, such as a chain of $L$ two-level systems, where $L$ is the width of the lattice. The state of $(1+1)D$ QCA then evolves iteratively as $\ket{\psi_{t+1}}=\mathcal{G}_{t}\ket{\psi_{t}}$, where $\mathcal{G}_t$ is a global-update operator defined via the application of local (unitary) gates $G_{k,t}$, $\forall k$. Each $G_{k,t}$ acts on a set of so-called control sites in the neighborhood of site $k$ in row $t$ and on the target site $k$ in row $t+1$, see an example in Fig.~\ref{fig:QCA_models}(b). The successive update of consecutive rows allows one to interpret the vertical dimension of the lattice [cf.~Fig.~\ref{fig:QCA_models}(a)] as an effective time dimension for the evolution of the one-dimensional system encoded in the QCA. Its (reduced) state, at time $t$, is given by the density matrix $\rho(t)=\Tr_{t}\left[\ket{\psi_{t}}\!\bra{\psi_{t}}\right]$, with $\Tr_{t}$ indicating trace over all sites but those in row $t$. Since $\mathcal{G}_t$ acts non-trivially only on two adjacent rows, the (reduced) dynamics of $\rho(t)$ is given by
\begin{align}
\rho({t+1}) =\Tr_{t+1}\left[\G_t \rho({t}) \otimes \ketbra{0}{0}_{t+1} \G^{\dag}_t\right] ~,
\label{eqn:qca_reduced_dynamics}
\end{align}
where $\ket{0}_{t+1}$ is the state of row $t+1$, initialized with all (target) sites in $\ket{\circ}$. In order not to overload the notation, we will use here the symbol $\G_t$ to indicate both the full unitary operator and its reduced form solely supported on row $t$ and $t+1$. 

As illustrated in Fig.~\ref{fig:QCA_models}(a), $(1+1)D$ QCA are structurally equivalent to layered QNNs composed of perceptrons \cite{Beer2020, Sharma2022}. The input layer of the QNN is the initial state of the one-dimensional system encoded in the QCA, while the remaining nodes of the network, i.e., those in the  hidden layers and in the output layer, are initialized in the {\it fiduciary} state $\ket{\circ}$. The (neural) nodes of two adjacent layers are connected by (unitary) perceptrons, which are implemented by the application of the gate $\G_t$ in the QCA. In the context of QNNs, the time dimension of the QCA represents a direction quantifying the progress of a computation, and the state at the output layer corresponds to the state $\rho(t)$ of the QCA, as defined in Eq.~\eqref{eqn:qca_reduced_dynamics}, at a chosen final time [cf.~Fig.~\ref{fig:QCA_models}(a)]. Since we consider here gates $G_{k,t}$ which are identical for all $t$ and $k$, our QCA in fact reproduces a recurrent QNN \cite{Broughton2020}.  

\noindent \textbf{Information processing through dissipative quantum dynamics.} As discussed above, $(1+1)D$ QCA, and thus also QNNs, encode a discrete-time dynamics [cf.~Eq.~\eqref{eqn:qca_reduced_dynamics}]. Here, we introduce a class of gates (perceptrons), $G_{k,t}$, for which, in a given limit of their parameters, this dynamics becomes equivalent to a continuous-time many-body open quantum system evolution. This is of potential interest for the learning of dissipative quantum dynamics -- with similarly highly structured perceptrons having proved important for learning closed, Hamiltonian dynamics \cite{Broughton2020}. However, this construction also enables the examination of the information processing of QNNs in terms of equivalent dynamical properties, see, e.g., Ref.~\cite{Verstraete2009}. 

The dissipative quantum dynamics we will consider is described by the so-called Lindblad generator
\cite{Lindblad1976}
\begin{equation}
\mathcal{L}[\rho]=-i[H,\rho]+\sum_\mu \left(J_\mu \rho J_\mu^\dagger -\frac{1}{2}\left\{J_\mu^\dagger J_\mu,\rho \right\}\right)\, .
\label{eq:Lindblad}
\end{equation}
Here the commutator term, which includes the Hamiltonian $H$, gives rise to a coherent quantum evolution, while the dissipator, which depends on the jump operators $J_\mu$, encodes a non-reversible (classical) process. For the sake of concreteness, we consider a QCA with gates $G_{k,t}$ acting on two control sites and a single target site, as is the case for the gate shown in Fig.~\ref{fig:QCA_models}(b). Generalizations to more control sites are straightforward. 

In order for the dynamics in Eq.~\eqref{eqn:qca_reduced_dynamics} to be (approximately) equivalent to the continuous-time dynamics generated by $\mathcal{L}$, we require that $\rho({t+1})\approx \rho(t) +\delta t  \mathcal{L}[\rho(t)]$, where $\delta t$ is a small --- as compared to the time-scales of the system ---  time increment. This can be achieved (see supplemental material \cite{SM}) via a global-update operator $\G_t$, composed of local gates [cf.~Fig.~\ref{fig:QCA_models}(b)]
\begin{align}
    G_{k, t} = \mathrm{SWAP}_{k}D_{k}U_{k-1,k} ~,
\label{eqn:local_gate_form}
\end{align}
applied in a right-to-left sweep gate-ordering. Here, $U_{k-1,k}=e^{-i \delta t h_{k-1,k}}$ (with the special case $U_{0,1} = \mathbb{I}$, i.e. open boundary conditions) unitarily evolves the control sites $k-1,k$ with the local Hamiltonian $h_{k-1,k}$. The operator $D_k$, given by 
\begin{align}
    D_{k} = \exp\left[i\theta\left(\sigma^+_{k} \otimes \sigma^-_{k} + \sigma^-_{k} \otimes \sigma^+_{k}\right)\right] ~,
\label{eqn:single_site_decay}
\end{align}
with $\sigma^-=(\sigma^+)^\dagger=\ket{\circ}\!\bra{\bullet}$, generates entanglement between the control site $(k,t)$ and the target site $(k,t+1)$ which results in dissipation in the form of local decay from state $\ket{\bullet}$ to state $\ket{\circ}$ at site $k$. The parameter $\theta$ needs to be small in order to approximate a continuous-time update. For small $\delta t$, we assume $\theta \approx \sqrt{\gamma \delta t}$, where $\gamma$ provides the decay rate. Finally, the operator $\mathrm{SWAP}_{k}$ performs a swap of site $(k,t)$ with site $(k,t+1)$, which is needed to advance the state of the system to the next row. Note that, in case of a fully coherent Hamiltonian evolution one has $D_k=\mathbb{I}$, and the target sites become redundant  \cite{Wiesner2009,Cirac2017,Arrighi2019,Farrelly2020,Hillberry2021}. 

In order to see that using the local gates in Eq.~\eqref{eqn:local_gate_form} gives rise to a continuous-time Lindblad dynamics, we consider Eq.~\eqref{eqn:qca_reduced_dynamics}, which in the present case reads \cite{SM},
\begin{align}
  \rho({t+1})=\sum_{m} \mathcal{K}_{m} U_{t} \rho(t) U_{t}^{\dag} \mathcal{K}_{m}^{\dagger} ~.
\label{eqn:rho_update}
\end{align}
Here, $m = (m_{1}, m_{2}, ..., m_{L})$, with $m_k=\circ,\bullet$ labelling the basis states. The operator $U_{t} = \prod_{k} U_{k-1,k}$ implements the full unitary update, while $\mathcal{K}_{m}=\prod_{k} K_{m_k, k}$, with $K_{m_{k}, k} = \bra{m_{k}}D_{k}\ket{\circ}$, are the (Kraus) operators associated with local decay. Recalling Eq.~\eqref{eqn:single_site_decay}, one has
\begin{align}
K_{\circ, k}= \mathbb{I}+(\cos(\theta) - 1)\, n_k  ~, \quad   K_{\bullet, k}= i\sin(\theta) \sigma^-_k\, ,
\end{align}
where $n=\ket{\bullet}\!\bra{\bullet}$. For small angles $\theta \approx \sqrt{\gamma \delta t }\ll 1$, we can expand the Kraus operators $\mathcal{K}_{m}$ in powers of $\delta t$. Up to first order in $\delta t$, in Eq.~\eqref{eqn:rho_update} one has at most a single $m_k=\bullet$. Since this can happen for all $k$, the sum of these contributions leads to a term equivalent to the first sum in Eq.~\eqref{eq:Lindblad}, acting on $U_t\rho(t) U^\dagger_t$. Expanding the term with all $m_k=\circ$ in Eq.~\eqref{eqn:rho_update} gives instead the second sum in Eq.~\eqref{eq:Lindblad} applied to $U_t\rho(t) U_t^\dagger$ as well as the contribution $U_t\rho(t)U_t^\dagger$ itself. Finally, expanding also the $U_{k-1,k}$ up to first order in $\delta t$, gives rise to a Lindblad operator which has the form of  Eq.~\eqref{eq:Lindblad}, with Hamiltonian $H=\sum_k h_{k-1,k}$ and jump operator $J_k=\sqrt{\gamma} \sigma_k^-$.

\noindent \textbf{Emergent collective many-body behavior.} Having shown that QNNs effectively implement open quantum dynamics, we exploit this connection to investigate emergent quantum many-body behavior in large-scale QNNs. To this end, we focus on a scenario in which critical behavior associated with an NEPT can be observed. We achieve this by considering a Hamiltonian with $h_{k,k+1} = \Omega \left(\sigma^{y}_{k}n_{k+1} + n_{k}\sigma^{y}_{k+1}\right)$ and $\sigma^y=-i \ket{\bullet}\!\bra{\circ}+{\rm h.c.}$, which establishes coherent oscillations between the two states of a given site only if at least one of the neighboring sites is in state $\ket{\bullet}$, e.g. $\ket{\circ\bullet}\leftrightarrow\ket{\bullet\bullet}$, a process which can be interpreted as quantum branching and coagulation.
The reason for this choice is that, in combination with local decay, it gives rise in the small $\delta t$ limit to the so-called quantum contact process (QCP) \cite{Carollo2019,Jo2021}. The ensuing dynamics displays an NEPT into an absorbing state (with all sites in $\ket{\circ}$) and emergent critical behavior influenced by quantum effects \cite{Carollo2019}. Due to universality this phase transition also emerges when $\delta t$ is not small, which is certainly the case in a generic QNN, where the gates $G_{k,t}$ implement discrete updates.

To systematically analyze this model, we reparametrize it as follows: $\theta =\arcsin(p_2)$, with $p_{2} = \sin^{2}\left(\sqrt{\gamma \delta t / 2}\right)$ and $\Omega \delta t=(1/\sqrt{2})\arcsin \sqrt{p_1}$. The two real parameters $p_{1/2} \in [0,1]$ can be interpreted as probabilities associated with the transitions $\ket{\circ\bullet}\leftrightarrow\ket{\bullet\bullet}$ due to the coherent branching/coagulation, and with the local decay $\ket{\bullet}\to\ket{\circ}$, respectively. The continous-time dynamics in Eq.~\eqref{eq:Lindblad} is then reproduced by Eq.~\eqref{eqn:qca_reduced_dynamics} when taking $p_{1},p_{2} \to 0$, for which $\delta t\to0$. 

\noindent \textbf{Mean-field analysis.} To investigate how quantum effects manifest in emergent behavior of large-scale QNNs, we study the stationary behavior of the associated QCA and look for signatures of NEPTs and critical behavior. We first derive the stationary phase diagram through a mean-field analysis, which allows us to gain insight on its overall structure \cite{Lesanovsky2019, Gillman2020, Gillman2021b} (see~\cite{SM} for details). 

\begin{figure}[t]
\centering
\includegraphics[width=1\linewidth]{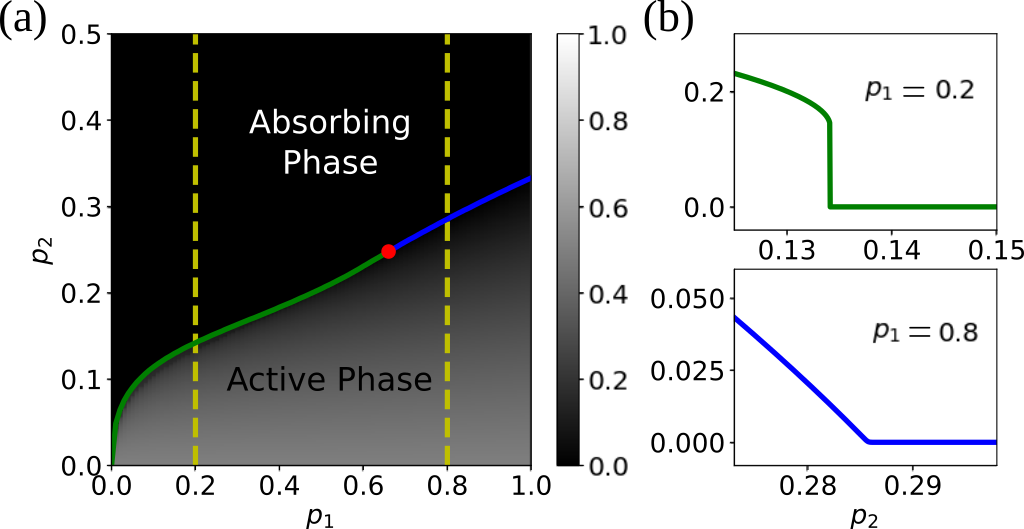}
\caption{\textbf{Mean-field phase diagram:} (a) Mean-field estimate for the stationary value of $\langle n \rangle$. 
To estimate the critical line, the gradient along each $p_{1}$-slice was calculated, and the point of maximum absolute gradient taken as the estimate for the critical $p_{2}$-value. The transition is estimated to be continuous when this gradient is below a threshold, set here to $10$ (the number of sampling points in $p_2$-direction is $2001$), and discontinuous otherwise. The separating ``transition point" is indicated by a red circle. 
(b) Two slices are shown for $p_{1} = 0.2, 0.8$. For these, the mean-field equations were iterated over $T=10000$ steps.
For $p_{1} = 0.2$, one can clearly observe a discontinuity in the stationary value of $\langle n\rangle$.}
\label{fig:mean_field_pd}
\end{figure}

In Fig.~\ref{fig:mean_field_pd}(a) we show the mean-field phase diagram. When $p_2$ is sufficiently large (large decay probability), the state $\rho(t)$ converges towards a stationary one with zero density of sites in the state $\ket{\bullet}$, i.e., $\langle n\rangle=0$. This phase is known as absorbing phase. On the other hand, for sufficiently small $p_2$, the stationary density $\langle n\rangle$ can be finite, $\langle n\rangle >0$. This establishes the existence of an active phase. The phase transition between these two regions appears discontinuous when $p_1 \lesssim 0.66$ and continuous otherwise, see Fig.~\ref{fig:mean_field_pd}(b). When $p_{1} = 1$, the mean-field equation for the density of occupied sites can be explicitly solved \cite{SM} giving solutions $\braket{n} = 0$ as well as,
\begin{align}
\braket{n}= \frac{3}{2} + \frac{1}{p_{2} - 1} ~.
\end{align}
This second solution vanishes continuously approaching $p_{2} \to 1/3$ from above. This behaviour is similar to that expected from a (mean-field) NEPT in the DP universality class, which is displayed by the CCP, i.e., the classical version of the contact process in which branching and coagulation occur as incoherent non-reversible processes. 

These findings show that a large QNN can indeed display emergent collective behavior akin to that of an NEPT. For $p_{1} = 1$, it appears that the QNN/QCA, as constructed here, exhibits classical collective behavior. However, reducing the value of $p_1$ changes the nature of the phase transition phenomenology, from a  continuous to a first-order one. A similar phenomenon was identified in an open quantum contact process model with competing classical (incoherent) and quantum (coherent) branching and coagulation \cite{Marcuzzi2016}. There, the discontinuous mean-field transition turns out to in fact be a continuous NEPT whose universal exponents are different from DP due to quantum effects \cite{Carollo2019}. This suggests that quantum fluctuations indeed can impact on the processing of information and concomitant collective effects in QNNs. 

\noindent \textbf{Tensor-network analysis.} To confirm that tuning $p_1$ indeed changes the NEPT, we employ (non-perturbative) TN methods. The approach taken here follows that of Ref.~\cite{Gillman2021b}, and directly approximates the evolution of $\rho(t)$ in a vectorised representation. We consider an initial state with all sites in $\ket{\bullet}$ and the evolution of the density $\langle n\rangle$ up to $T=100$ time-steps. 
\begin{figure}[t]
\centering
\includegraphics[width=1\linewidth]{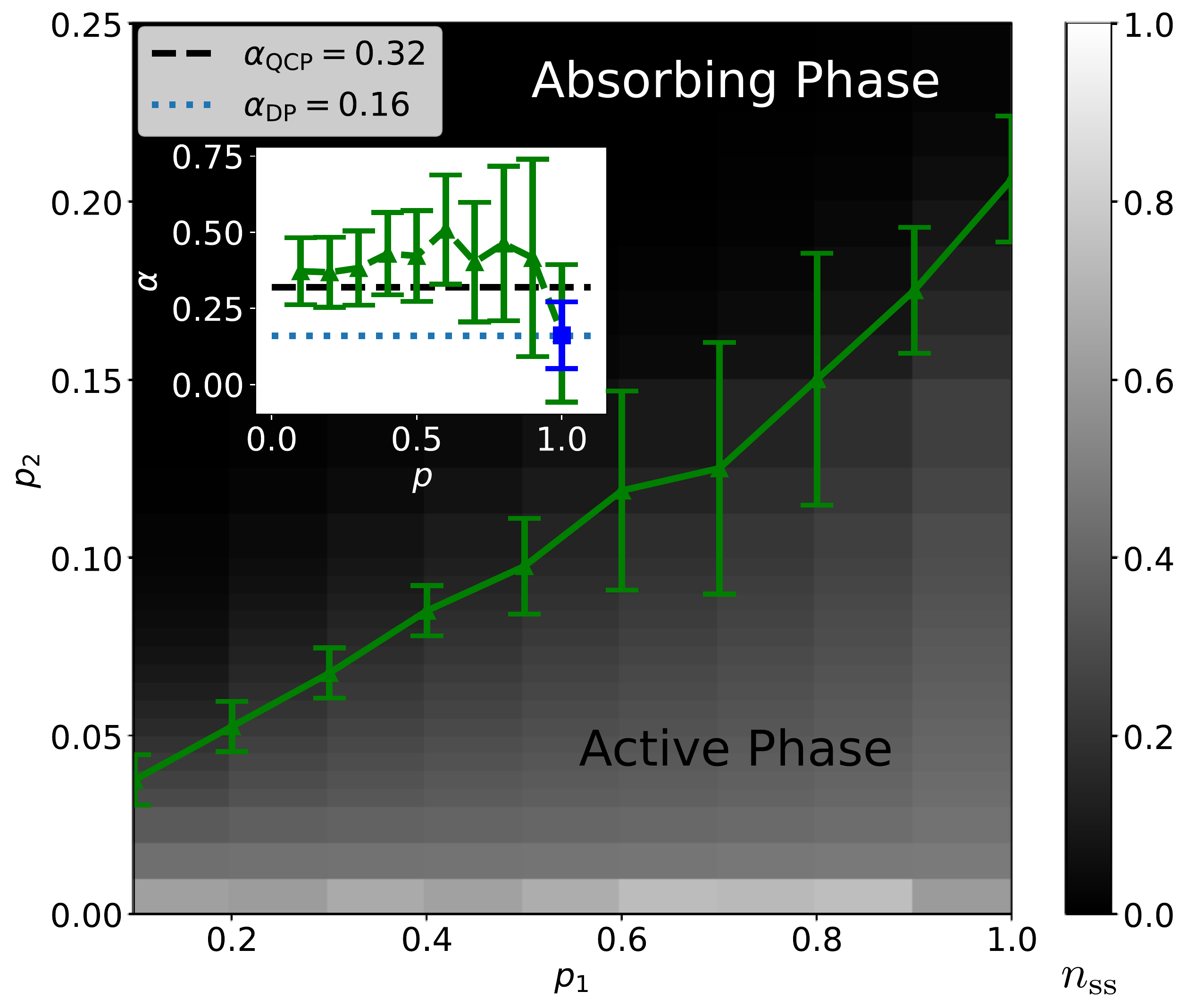}
\caption{\textbf{Tensor-network phase diagram:} The density $\langle n\rangle$ is estimated by evolving $\rho(t)$ over $T=100$ steps, with a bond-dimension of $\chi=512$, lattice size of $L=128$ and starting from a fully occupied state. The critical line shown in green (triangles) is estimated by taking the average of two estimates, one using linear fits of $\langle n(t)\rangle$ in a log-log scale, and another using an effective exponent $\alpha(t)$, see \cite{SM} for details. The error bars combine the two errors on these estimated values of $p_{2}$ (each induced by the resolution of the parameter grid). The inset shows the corresponding critical exponents (green triangles), obtained by averaging $\alpha(t)$ over $t \in [80, 100]$. Errors occurring on each effective exponent separately are indicated by the error bars and involve contributions from finite-sizes, finite bond-dimensions and finite parameter grid resolutions \cite{SM}. As can be seen, errors tend to increase with $p_{1}$. To gain a higher accuracy estimate the exponent for $p_{1} = 1$, the procedure was repeated with $L=256$ (blue square in the inset). The dotted (dashed) line shows the exponent $\alpha$ expected for the DP (QCP) universality class.}
\label{fig:tn_pd}
\end{figure}

The first task is to estimate the position of the critical line separating the active phase and the absorbing phase. To this end, we take two different approaches, both exploiting that $\langle n (t)\rangle$ obeys a power-law behavior in time on the critical line.  This means that, at criticality, $\langle n(t)\rangle\sim t^{-\alpha}$, where $\alpha$ is a dynamical exponent characterizing the universal behavior of the model. Details on the methods employed are given in~\cite{SM}. 

As shown in Fig.~\ref{fig:tn_pd}, the TN phase diagram qualitatively agrees with the mean-field one, although the active phase seems to extend to higher values of $p_{2}$ in the latter case. 
In order to gain insights on the critical behavior of the model, we analyze the critical exponent $\alpha$ by investigating the behavior of the effective exponent $\alpha(t)=-\log_2 \langle n(2t)\rangle/\langle n(t)\rangle$. The latter becomes constant for exact power laws, in which case it further provides an estimate for $\alpha$ (see Ref.~\cite{SM} for details). 
 
As shown in the inset of Fig.~\ref{fig:tn_pd}, while the precise estimation of $\alpha$ and associated errors is challenging, for $p_{1} \le 0.5$, the obtained value of the exponent always lies far from the $1D$ DP value $\alpha_{\mathrm{DP}} \approx  0.16$, and is in fact close to the one of the $1D$ QCP,  $\alpha_{\mathrm{QCP}} = 0.32$ \cite{Carollo2019}. This suggests that, despite being far from the continuous-time limit, the dynamics displays the same critical behaviour as the QCP and, unlike the CCP, does not fall into the DP universality class. However, as shown in the inset, errors -- particularly due to finite-size effects -- increase with $p_{1}$ and determining the universality class in this regime becomes more and more challenging. It is therefore difficult to reliably estimate a particular location where the universality class changes from that of the QCP to DP. This issue is compounded by the suggestion that, in fact, due to the competition between quantum and classical effects, the dynamical exponent may even vary continuously \cite{Jo2021}. As such, we here consider the extreme point $p_{1} = 1$ and run a finer parameter scan with a larger system size to reduce errors. The resulting estimate for the critical exponent lies very close to that of $1D$ DP [cf.~inset of Fig.~\ref{fig:tn_pd}], indicating that indeed the universality class must change at some point, as suggested by the mean-field analysis. 

\noindent \textbf{Conclusion.} We have established a connection between $(1+1)D$ QCA, emerging as natural generalizations of CCA for the study of quantum nonequilibrium processes, and perceptron-based QNNs. We have shown how these platforms can encode continuous-time open quantum dynamics in limiting cases, and have exploited this to investigate the emergence of quantum critical behavior in large-scale QNNs, i.e., deep QNNs with many intra-layer nodes. Our results show that the processing of information within  QNNs can be understood as an open quantum many-body dynamics and that quantum effects indeed influence their large-scale collective behavior.

\textbf{Acknowledgments.} We are grateful to Markus M\"uller and Mario Boneberg for fruitful discussions. We acknowledge support from EPSRC [Grant No. EP/V031201/1], from the ``Wissenschaftler R\"{u}ckkehrprogramm GSO/CZS" of the Carl-Zeiss-Stiftung and the German Scholars Organization e.V., through The Leverhulme Trust [Grant No. RPG-2018-181], and the Deutsche Forschungsgemeinschaft through Grants No. 449905436, 435696605 as well as through the Research Unit FOR 5413/1, Grant No. 465199066. We are grateful for access to the University of Nottingham's Augusta HPC service.

\bibliographystyle{apsrev4-1}
\bibliography{QCA_bib}

\onecolumngrid
\newpage

\pagebreak
\widetext

\begin{center}
\textbf{\large Supplemental Material}
\end{center}

\setcounter{section}{0}
\setcounter{equation}{0}
\setcounter{figure}{0}
\setcounter{table}{0}
\setcounter{page}{1}
\makeatletter

\renewcommand\thesection{S\arabic{section}}
\renewcommand{\theequation}{S\arabic{equation}}
\renewcommand{\thefigure}{S\arabic{figure}}
\renewcommand{\thetable}{S\arabic{table}}
\renewcommand{\bibnumfmt}[1]{[S#1]}
\section{Construction of QCA Gates for Lindblad Dynamics}

In this section, we show how discrete-time $(1+1)D$ QCA with local gates given in Eq.~\eqref{eqn:local_gate_form} can provide a first-order approximation of a given continuous-time Lindblad dynamics of a one-dimensional quantum system. As done in the main text, we focus on the case of single-site decay ($\ket{\bullet}\to\ket{\circ}$) dissipation and nearest-neighbour Hamiltonian.

\subsection{Local gate}

First, we recall the definition of the local gate considered in the main text,
\begin{align}
    G_{ k, t} = \mathrm{SWAP}_{k}D_{k}U_{k-1,k} ~,
\label{eqn_sm:local_gate_form}
\end{align}
which acts on the target site $k$ on row $t+1$ and control sites $k-1,k$ on row $t$. 
Here,
\begin{align}
    D_{k} = \exp\left[i\theta\left(\sigma^+_{k} \otimes \sigma^-_{k} + \sigma^-_{k} \otimes \sigma^+_{k}\right)\right] ~
\label{eqn_sm:single_site_decay}
\end{align}
entangles control site $k$ and target site $k$ while
\begin{align}
\text{SWAP}_{k} &= n_{k}\otimes n_{k} + \nb_{k}\otimes \nb_{k} + \sigma^+_{k}\otimes \sigma^-_{k} + \sigma^-_{k}\otimes\sigma^+_{k} ~, 
  \label{eqn_sm:swap_gate}
\end{align}
with $\nb=\mathbb{I}-n$, swaps them. 
The unitary $U_{k-1, k}$ solely acts on control sites $k-1,k$ and is defined through the nearest-neighbor Hamiltonian term $h_{k-1,k}$ as
\begin{align}
U_{k-1, k} &= \exp\left[ -i \delta t  h_{k-1,k}\right] ~,
\end{align}
with the special case $U_{0,1} = \mathbb{I}$.

\subsection{Global-update operator}
Through the above local unitary gates, we can define a global update, assuming a ``right-to-left" local-gate ordering, as 
\begin{equation}
  \mathcal{G}_{t} = \prod_{k=1}^{N} G_{k,t} = G_{1,t}G_{2,t}\dots G_{k,t}\dots G_{L-1,t}G_{L,t} .
\end{equation}
Exploiting the form of the gate $G_{k,t}$ above, we can write
\begin{align}
  \mathcal{G}_{t} &= \prod_{k=1}^{N} G_{k,t} ~ ,  \\ 
   &=\mathrm{SWAP}_1 D_1 \prod_{k=2}^{N}\left(\mathrm{SWAP}_{k} D_{k} U_{k-1,k}\right) ~, \nonumber \\
  &=\mathrm{SWAP}_1 D_1 U_{0,1} \mathrm{SWAP}_2 D_2 U_{1,2} \mathrm{SWAP}_3 D_3 U_{2,3} ... \mathrm{SWAP}_N \bar{D}_N U_{N-1,N} \\
   &=\mathrm{SWAP}_1 D_1  \mathrm{SWAP}_2 D_2 \mathrm{SWAP}_3 D_3 ... \mathrm{SWAP}_N D_N \left(U_{0,1}  U_{1,2} U_{2,3} ... U_{N-1,N} \right) \\
  &= \left(\prod_{k=1}^{N} \mathrm{SWAP}_k D_k\right)\left(\prod_{k=1}^{N} U_{k-1,k}\right) ~.
\label{eqn_sm:global_update_factorisation}
\end{align}
In the third line, we used that the unitary operators $U_{k-1,k}$ commute with all $\mathrm{SWAP}_{k}$ and $D_{k}$ to their right.

In the case that $D_{k}$ and $\mathrm{SWAP}_{k}$ commute, as is the case, e.g., for single-site decay, then $\mathcal{G}_{t}$ further factorises as,
\begin{align}
  \mathcal{G}_{t} &= \left(\prod_k \mathrm{SWAP}_k\right)\left(\prod_k D_k\right)\left(\prod_k U_{k-1,k}\right) ~.
\label{eqn_sm:global_update_factorisation_three}
\end{align}

From now one, we will make the row-support of operators explicit. We will then write the global update as
\begin{eqnarray}
  \mathcal{G}_{t,t+1}=S_{t,t+1} D_{t,t+1} (U_t \otimes \mathbb{I}_{t+1}) ~ ,
\end{eqnarray}
where $S_{t,t+1}=\prod_k \mathrm{SWAP}_k$ swaps neighboring qubits of adjacent rows and $D_{t,t+1}=\prod_k D_{k}$ entangles opposite qubits on the two rows. In the above formula, we also made explicit that the unitary operator $U_t$ solely acts on row $t$. The rationale behind this decomposition of $\mathcal{G}_{t,t+1}$ is that $D_{t,t+1}$ introduces dissipative processes, $U_t$ implements the Hamiltonian part of the  evolution on row $t$ and $S_{t,t+1}$ advances the state from row $t$ to row $t+1$.\\

We are now in the position to write the time evolution under this global update gate. First, defining $\tilde{\rho}(t) = U_{t} \rho(t) U_{t}^{\dag}$, we can write the update as [we also write explicitly the row index in $\rho(t)$ inside the trace]:
\begin{eqnarray}
  \rho(t+1)&=&\mathrm{Tr}_{t+1} \left[S_{t,t+1} D_{t,t+1}  (U_t \otimes \mathbb{I}_{t+1}) \rho(t)_t\otimes|0\rangle \langle 0 |_{t+1} (U^\dagger_t \otimes \mathbb{I}_{t+1}) D_{t,t+1}^\dagger S^\dagger_{t,t+1} \right]\\
 &=&\mathrm{Tr}_{t+1} \left[S_{t,t+1} D_{t,t+1} \tilde{\rho}(t)_t\otimes|0\rangle \langle 0 |_{t+1} D_{t,t+1} S_{t,t+1} \right]\, , 
 \end{eqnarray}
where we also used that $D_{t,t+1}=D_{t,t+1}^\dagger$ and $S_{t,t+1}=S_{t,t+1}^\dagger$. Then, exploiting the commutativity between $S_{t,t+1}$ and $D_{t,t+1}$, we have
\begin{eqnarray}
  \rho(t+1)&=&\mathrm{Tr}_{t+1} \left[ D_{t,t+1} S_{t,t+1} \tilde{\rho}(t)_t\otimes|0\rangle \langle 0 |_{t+1} S_{t,t+1} D_{t,t+1} \right]\, .
\end{eqnarray}
We now apply the swap operation on the state and the auxiliary row of target sites in $\ket{\circ}$ which leads to 
\begin{eqnarray}
  \rho(t+1)&=&\mathrm{Tr}_{t+1} \left[ D_{t,t+1} |0\rangle \langle 0 |_{t} \otimes\tilde{\rho}(t)_{t+1} D_{t,t+1} \right]\, .
\end{eqnarray}
We can finally perform the trace over row $t$. To this end, we introduce the basis states $\ket{m}$, where $m=(m_1,m_2,\dots, m_L)$ and $m_k=\circ,\bullet$ specify the state of the $k$th site in row $t$. We thus have 
\begin{eqnarray}
  \rho(t+1)&=&\mathrm{Tr}_{t+1} \left[ D_{t,t+1} |0\rangle \langle 0 |_{t} \otimes\tilde{\rho}(t)_{t+1} D_{t,t+1} \right]\\
 &=&\sum_{m\in[\circ,\bullet]^L}\bra{m} \left[ D_{t,t+1} |0\rangle \langle 0 |_{t} \otimes\tilde{\rho}(t)_{t+1} D_{t,t+1} \right]\ket{m}\\
 &=&\sum_{m\in[\circ,\bullet]^L} \bra{m} D_{t,t+1} |0\rangle_{t} \tilde{\rho}(t)_{t+1} \bra{0} D_{t,t+1}^\dagger\ket{m}_{t}\\ &=&\sum_{m\in[\circ,\bullet]^L} \left[\prod_{k} K_{m_k, k}\right] \tilde{\rho}(t) \left[\prod_{k} K^\dagger_{m_k, k}\right] ~.
\label{eqn_sm:rho_update}
\end{eqnarray}
In the final line, we introduced the single-particle Kraus operators, $K_{m_k, k} = \braket{m_{k}|D_{k}|\circ}$, labelled by $m_{k} = \circ,\bullet$ and the site $k$. These come into play after the decomposition of $\bra{m} D_{t,t+1} |0\rangle_t$ into local terms, as shown in the following equation
\begin{align}
    \bra{m} D_{t, t+1} \ket{0}_{t} &= \bra{m_{1}, m_{2}, ..., m_{L}} D_{t, t+1} \ket{\circ_{1}\circ_{2}...\circ_{L}}_{t}  \\
    &= \bra{m_{1} m_{2} ...m_{L}} \prod_k \exp\left[i\theta\left(\sigma^+_{k} \otimes \sigma^-_{k} + \sigma^-_{k} \otimes \sigma^+_{k}\right)\right] \ket{\circ_{1}\circ_{2}...\circ_{L}}_{t} ~, \\
    &=  \prod_k \bra{m_{k}}\exp\left[i\theta\left(\sigma^+_{k} \otimes \sigma^-_{k} + \sigma^-_{k} \otimes \sigma^+_{k}\right)\right] \ket{\circ_{k}}_{t} \\
    &= \prod_{k} K_{m_{k}, k} ~.
\end{align}
Note that in Eq.~\eqref{eqn_sm:rho_update}, we removed the subscript $t$ from $\tilde{\rho}(t)$ since there is no more ambiguity on the rows, given that the trace has already been taken. 

The single-site Kraus operator $K_{m_k,k}$ have the form,
\begin{eqnarray}
  K_{\circ, k}&=& \mathbb{I}+(\cos\theta - 1)\, n_k \, ,\nonumber\\
  K_{\bullet, k}&=& i\sin\theta\, \sigma^-_k,
\label{eqn_sm:single_particle_kraus}
\end{eqnarray}
as it is readily found by noting that
\begin{eqnarray}
  \exp\left[i\theta\left(\sigma^+_k \, \otimes \sigma^-_k + \sigma^-_k \otimes  \sigma^+_k\right)\right]&=&n_k\otimes n_k+ (1-n_k)\otimes (1-n_k)\\
   &&+\cos\theta \left[(1-n_k)\otimes n_k+n_k \otimes (1-n_k)\right]+i\sin\theta\left[\sigma^+_k \otimes \sigma^-_k + \sigma^-_k \otimes \sigma^+_k\right],\nonumber
\end{eqnarray}
which yields
\begin{eqnarray}
  &_t\langle \circ |\exp\left[i\theta\left(\sigma^+_k\otimes  \, \sigma^-_k + \sigma^-_k \otimes  \sigma^+_k\right)\right]|\circ\rangle_t=\mathbb{I}+\cos\theta\, n_k\\
  &_t\langle \bullet |\exp\left[i\theta\left(\sigma^+_k  \otimes \sigma^-_k + \sigma^-_k \otimes \sigma^+_k\right)\right]|\circ\rangle_t=i\sin\theta\, \sigma^-_k.
\end{eqnarray}

Defining $\mathcal{K}_{m}= \prod_{k} K_{m_{k}, k} $ the reduced evolution of $\rho(t)$ can thus be written as
\begin{align}
    \rho(t+1)=\sum_{m\in[\circ,\bullet]^L} \mathcal{K}_{m} U_{t} \rho(t) U_{t}^{\dag} \mathcal{K}_{m}^{\dagger} ~.
    \label{eqn_sm:reduced_evo_final}
\end{align}
Note that this includes the action of all products of the single-particle Kraus operators.

\subsection{Equivalence with Lindblad dynamics}
We now can show that Eq.~\eqref{eqn_sm:reduced_evo_final} can approximate a Lindblad dynamics up to first order in the discretization  time-step $\delta t$. 
To this end, we start recalling that under a continuous-time Lindblad dynamics the evolution of a density matrix $\rho(t)$ is implemented by the quantum master equation \begin{equation}
\dot{\rho}(t)=\mathcal{L}[\rho(t)]
\label{eq:QME}
\end{equation}
with 
\begin{align}
\mathcal{L}[\rho]=-i[H,\rho]+\mathcal{D}[\rho]\, .
\label{eqn_sm:lindblad}
\end{align}
The first term in Eq.~\eqref{eqn_sm:lindblad} represents the evolution due to the  system Hamiltonian, i.e., via the coherent part of the dynamics. The term $\mathcal{D}$ contains information about the effects of the environment, or of stochasticity in general, on the system. This has the general form,
\begin{align}
\mathcal{D}[X]=\sum_\mu \left(J_\mu XJ_\mu^\dagger -\frac{1}{2}\left\{X, J_\mu^\dagger J_\mu\right\}\right)\, ,    
\end{align}
with $\{A,B\}=AB+BA$.

The dynamics under Eq.~\eqref{eq:QME} can be approximated, up to first order via the update relation 
\begin{align}
    \rho(t+1) \approx \rho(t) + \delta t \mathcal{L}\left[\rho(t)\right] ~.
\label{eqn_sm:first_order_lind}
\end{align}
Our goal is now to show that the QCA update in Eq.~\eqref{eqn_sm:reduced_evo_final} exactly corresponds to the above approximation. 

Let us now consider the global QCA update \eqref{eqn_sm:reduced_evo_final}, resulting from the local gate \eqref{eqn_sm:local_gate_form}.  Recall the form of the local unitary,
\begin{align}
U_{k,k+1} = \exp\left[- i \delta t  h_{k,k+1}\right] ~.
\end{align}
This defines a global unitary,
\begin{align}
    U_{t} &= \prod_{k} U_{k-1,k} ~, \nonumber \\
    &= \prod_{k}\exp\left[- i \delta t  h_{k,k+1}\right] ~.
\end{align}
To first order in $\delta t $, this is equivalent to the unitary,
\begin{align}
    \tilde{U}_{t} = \exp\left[- i \delta t  H\right]~,
\end{align}
where $H = \sum_{k} h_{k,k+1}$. Therefore, with this choice,
\begin{align}
    \tilde{\rho}(t) &= U_{t} \rho(t) U_{t}^{\dag} ~, \\
    &\approx \exp\left[- i \delta t  H\right] \rho(t) \exp\left[i \delta t  H\right] ~, \\
 &\approx  \rho(t)  - i  \delta t \left[H, \rho(t)\right] ~,
\end{align}
where the $``\approx"$ indicates here equivalence when neglecting terms greater than first-order in $\delta t$.

The global update Eq. \eqref{eqn_sm:reduced_evo_final} then has the form,
\begin{align}
\rho(t+1) \approx\sum_{m} \mathcal{K}_{m} \rho(t) \mathcal{K}_{m}^{\dagger} - i  \delta t \sum_{m} \mathcal{K}_{m} \left[H, \rho(t)\right] \mathcal{K}_{m}^{\dagger} ~.
\end{align}
Now, considering the definition of the Kraus operators in Eq.~\eqref{eqn_sm:single_particle_kraus}, and setting $\theta = \sqrt{\gamma \delta t}$, we also have,
\begin{align}
K_{\circ, k}&= \mathbb{I}+(\cos\theta - 1)\, n_k \approx \mathbb{I} - \frac{1}{2}\gamma \delta t n_{k} \, , \nonumber\\
K_{\bullet, k}&= i\sin\theta\, \sigma^-_k \approx i \sqrt{\gamma \delta t} \sigma^-_k\, .
\end{align}
Therefore,
\begin{align}
\rho(t+1) &\approx \sum_{m} \mathcal{K}_{m} \rho(t) \mathcal{K}_{m}^{\dagger} - i  \delta t \left[H, \rho(t)\right] \\
&\approx \rho(t) - \frac{1}{2} \gamma \delta t \sum_{k=1}^{L} \lbrace n_{k} , \rho(t) \rbrace + \gamma \delta t \sum_{k=1}^{L} \sigma^{-}_{k} \rho(t) \sigma^{+}_{k} - i  \delta t  \left[H, \rho(t)\right] \\
&= \rho(t) - i  \delta t  \left[H, \rho(t)\right] + \gamma \delta t \sum_{k=1}^{L}\left( \sigma^{-}_{k} \rho(t) \sigma^{+}_{k} - \frac{1}{2}  \lbrace n_{k} , \rho(t) \rbrace \right) ~,
\end{align}
which exactly corresponds to the desired approximation in the case of local losses with rate $\gamma$. 

\section{Continuous-Time QCP Limit}
In this section, we now present the continuous-time quantum contact process (QCP) model and discuss its implementation via QCA. 

\subsection{Continuous-time QCP}

In the QCP, the fundamental dynamical process is coherent (quantum) branching. This can be implemented in a Lindblad form by choosing
\begin{align}
H=\Omega \sum_{k=1}^{L-1}\left( \sigma_k^y n_{k+1}+ n_{k}\sigma^y_{k+1}\right) ~.
\label{eqn:H}
\end{align}
This Hamiltonian implements both a branching and a coagulation process at the same coherent rate, $\Omega$, via constrained Rabi oscillations at site $k$ that can only occur if the neighbors of $k$ are not simultaneously in the empty state.

For a system of size $L$, the dissipative term $\mathcal{D}$ is defined as,
\begin{align}
\mathcal{D}[\rho]=\gamma \sum_{k=1}^L\left(\sigma_k^{-}\rho\sigma_k^{+}-\frac{1}{2}\left\{n_{k},\rho\right\}\right)\, .
\end{align}

\subsection{QCA gate for the QCP}
A QCA containing the QCP as a limit can be constructed by choosing the local gate reported in Eq.~\eqref{eqn_sm:local_gate_form} with unitary implementing the Hamiltonian dynamics given by 
\begin{align}
U_{k-1, k} &= \exp\left[ -i \delta t  h_{k-1,k}\right] =  \exp\left[ -i \delta t \Omega \left(\sigma_{k}^{y} n_{k+1} + n_{k}\sigma^{y}_{k+1}\right)\right]~.
\end{align}

\begin{figure}[t]
\centering
\includegraphics[width=0.5\linewidth]{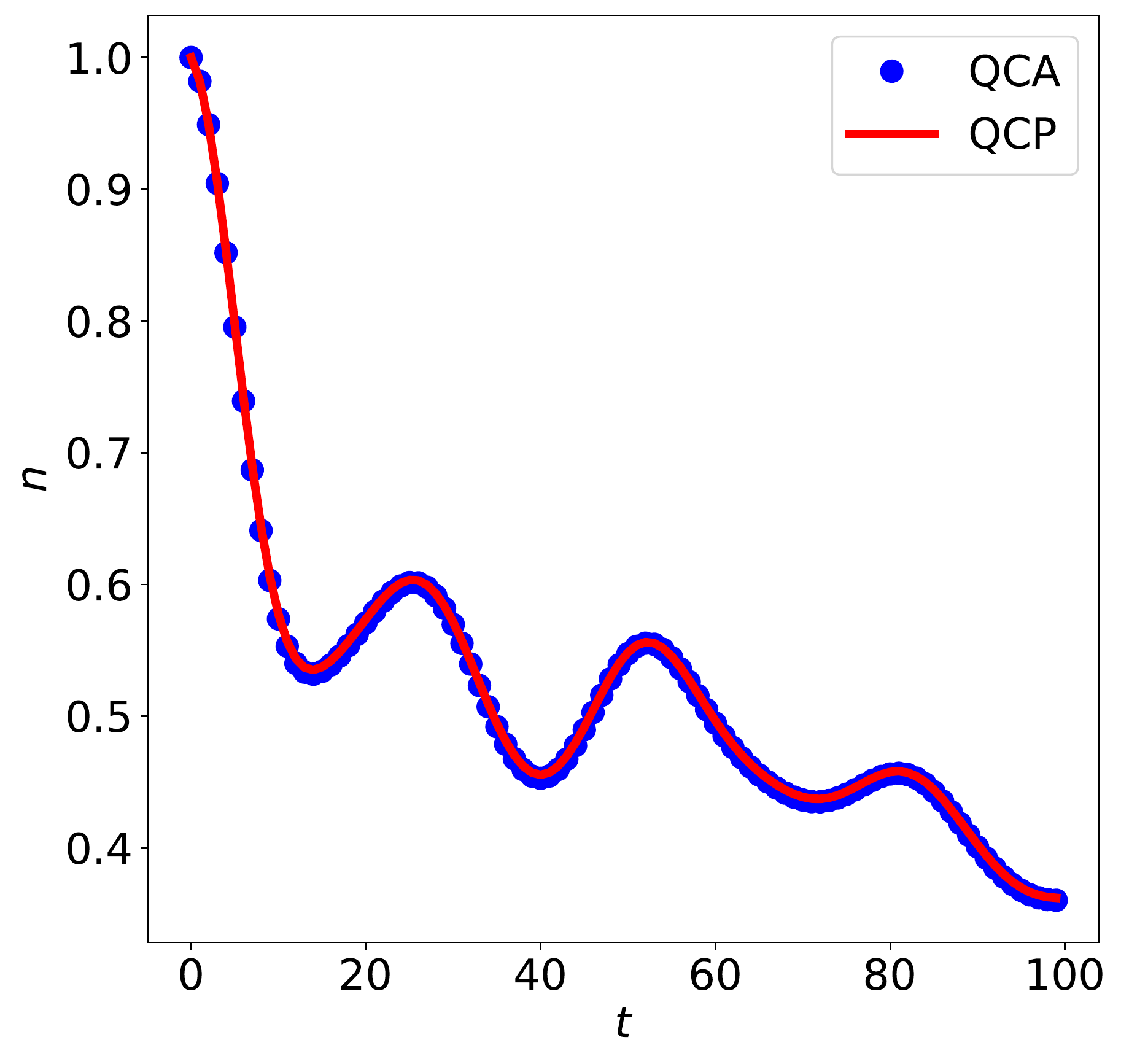}
\caption{The continuous time limit of the QCA. The expected density of the QCA is shown for $p_{1} = 0.006597$ and $p_{2} = 0.004991672$ using a right-to-left sweep ordering and averaged over all $L=4$ sites. Time is given in units of $\gamma^{-1}$. In terms of the QCP parameters, $p_{1} = \sin^{2}\left(\sqrt{2} \delta t \Omega\right)$ and $p_{2} = \sin^{2}\left(\sqrt{\gamma \delta t / 2}\right)$. This was simulated using TN methods and $\chi = 16$ . The QCP uses $\gamma \delta t = 0.01$ and a second-order Trotter scheme and bond-dimension $\chi = 16$, see \cite{Gillman2019} for a discussion of the method.}
\label{fig:qca_continous_time_limit}
\end{figure}

Setting $\theta = \sqrt{\gamma \delta t / 2}$, then taking $\delta t \to 0$, which implies $\gamma \delta t \ll 1$ and $\Omega \delta t \ll 1$, as shown before, the reduced dynamics of a row of the QCA agrees with the Lindblad evolution of the QCP up to first order in $\delta t$. This is illustrated numerically for $L=4$ by taking  $\Omega/\gamma = 5.75, \gamma \delta t = 0.01$ in Fig.~\ref{fig:qca_continous_time_limit}. 

\section{Mean-Field Theory for QCA with QCP gate}

In this section we consider a mean-field approximation of the reduced $1D$ dynamics of the state $\rho(t)$. The mean-field method we apply here follows similar prescriptions for $(1+1)D$ QCA taken previously \cite{Lesanovsky2019}. In this approach, one basically assumes that the state of each row is always in product form. This means that one has $\rho(t)=\otimes_{k=1}^L \rho^{(1)}(t)$, where $\rho^{(1)}(t)$ is the single-site reduced density matrix in the mean-field approximation. The evolution of the latter is found through the following iterative equation involving only two control sites and a target one
\begin{equation}
    \rho^{(1)}(t+1)={\rm Tr}_{t+1}\left[G \left(\rho^{(1)}(t)\otimes\rho^{(1)}(t)\right)_t\otimes \ket{\circ}\!\bra{\circ}_{t+1}G^\dagger \right]
\end{equation}
Here, $G$ is equal to gate $G_{k,t}$ and solely acts on the considered two control and one target sites.

The single-site density matrix, $\rho^{(1)}(t)$, can be alternatively specified at anytime by three expectation values, $\braket{n(t)}, \braket{\sigma^{x}(t)}$ and $\braket{\sigma^{y}(t)}$, where $\sigma_x=\sigma^++\sigma^-$ and $\sigma_y=-i\sigma^++i\sigma^-$. Because of this, we can monitor the evolution of $\rho^{(1)}(t)$ by simply focussing on the update rules for the expectation values. These can be computed as 
\begin{align}
\braket{n(t+1)} &= \left(1-p_{2}\right)\left[p_{1}\braket{n(t)}^{2} + \left(\frac{\sqrt{2}}{2}\sqrt{p_{1}}\sqrt{1-p_{1}}\braket{\sigma^{x}(t)} - \frac{p_{1}}{2} - 1\right)\braket{n(t)} - \frac{p_{1}}{8}\left(\braket{\sigma^{x}(t)}^{2}  + \braket{\sigma^{y}(t)}^{2}  \right) \right] ~. \\
\braket{\sigma^{y}(t+1)} &= \sqrt{1-p_{2}} \braket{\sigma^{y}(t)}\left( \braket{n(t)} \left[ 1 - \sqrt{1-p_{1}} \right] + \frac{\sqrt{2}}{2}\sqrt{p_{1}}\braket{\sigma^{x}(t)} + \sqrt{1-p_{1}} \right) ~. \\
\braket{\sigma^{x}(t+1)} &= \sqrt{1-p_{2}}\left( 2\sqrt{2}\sqrt{p_{1}}\sqrt{1-p_{1}} \braket{n(t)}^{2} - \left[2\sqrt{2}\sqrt{p_{1}}\sqrt{1-p_{1}} + \left(2p_{1} + \sqrt{1-p_{1}} - 1\right)\braket{\sigma^{x}(t)}\right] \braket{n(t)}\right) +  \nonumber \\
&~+ \sqrt{1-p_{2}}\left( \sqrt{1-p_{1}}\braket{\sigma^{x}(t)} - \frac{\sqrt{2}}{4}\sqrt{p_{1}}\left(\sqrt{1-p_{1}} - 1\right)\braket{\sigma^{x}(t)}^{2}\right) + \nonumber \\
& ~ - \frac{\sqrt{2}}{4}\sqrt{1-p_{2}}\sqrt{p_{1}}\left(\sqrt{1-p_{1}} + 1\right) \braket{\sigma^{y}(t)}^{2} ~ .
\end{align}

The second equation indicates that if $\braket{\sigma^{y}}$ is initialised as zero, it will remain so throughout. The two remaining equations then simplify to,
\begin{align}
\braket{n(t+1)} = \left(1-p_{2}\right)\left[p_{1}\braket{n(t)}^{2} + \left(\frac{\sqrt{2}}{2}\sqrt{p_{1}}\sqrt{1-p_{1}}\braket{\sigma^{x}({t})} - \frac{p_{1}}{2} - 1\right)\braket{n(t)}- \frac{p_{1}}{8}\left(\braket{\sigma^{x}(t)}^{2}\right) \right] ~.
\end{align}
\begin{align}
\braket{\sigma^{x}(t+1)} &= \sqrt{1-p_{2}}\left( 2\sqrt{2}\sqrt{p_{1}}\sqrt{1-p_{1}} \braket{n({t})}^{2} - \left[2\sqrt{2}\sqrt{p_{1}}\sqrt{1-p_{1}} + \left(2p_{1} + \sqrt{1-p_{1}} - 1\right)\braket{\sigma^{x}({t})}\right] \braket{n({t})} \right) +  \nonumber \\
&~+ \sqrt{1-p_{2}}\left( \sqrt{1-p_{1}}\braket{\sigma^{x}({t})} - \frac{\sqrt{2}}{4}\sqrt{p_{1}}\left(\sqrt{1-p_{1}} - 1\right)\braket{\sigma^{x}({t})}^{2}\right)  ~ .
\end{align}
These can be be iterated or solved directly [by setting $\braket{n({t+1})} = \braket{n({t})}$ and similarly for $\braket{\sigma^{x}}$] to approximate the steady-state density and hence produce a mean-field phase diagram.

In the special case $p_{1} = 1$ the nonequilibrium phase transition can be seen explicitly at the mean-field level, as the second equation above reduces to,
\begin{align}
\braket{\sigma^{x}({t}+1)} &= - \sqrt{1-p_{2}}\braket{\sigma^{x}({t})}\braket{n({t})} + \sqrt{1-p_{2}}\left(\frac{\sqrt{2}}{4}\braket{\sigma^{x}({t})}^{2}\right)  ~ .
\end{align}

Therefore, for this case, if $\braket{\sigma^{x}({t=0})} = 0$ then also  $\braket{\sigma^{x}({t})}=0$,  for all $t$. The equation for the density then becomes 
\begin{align}
\braket{n(t+1)} = \left(1-p_{2}\right)\left[\braket{n(t)}^{2} -\frac{3}{2}\braket{n(t)}\right] ~.
\end{align}
This has the stationary solutions $\braket{n} = 0$ and, 
\begin{align}
\braket{n}= \left(\frac{3}{2} + \frac{1}{p_{2} - 1}\right) ~.
\end{align}
This second solution vanishes continuously with $p_{2} \to 1/3$ from above.

\section{Estimation of the critical line and of the critical (dynamical) exponent}

In this section, we provide details on the estimation of the critical line, separating the absorbing phase from the active one, from the numerical results obtained with tensor networks. We further classify the nonequilibrium phase transition in the different parameter regions by analyzing and estimating the critical power-law exponent $\alpha$, governing the decay of the density of sites in $\bullet$ at criticality. 

\subsection{Simulation method and parameter choices}

As discussed in the main text, to estimate the critical exponents, the time evolution of the density $\langle n\rangle$ of the chosen QCA was first approximated using the tensor-network (TN) method described in Ref.~\cite{Gillman2021b}. In that approach, the reduced state $\rho(t)$ of the one-dimensional system is vectorised via a mapping to the ``doubled-space", $\rho({t}) \to |\rho ({t})\rangle\hspace{-2pt}\rangle$. The resulting vector is then represented as matrix product state (MPS) tensor  network \cite{Eisert2013}. The corresponding evolution equation can be written as, $|\rho(t+1)\rangle\hspace{-2pt}\rangle= \Lambda |\rho(t)\rangle\hspace{-2pt}\rangle$, where the linearised evolution operator, $\Lambda$, can be represented as a matrix product operator (MPO) tensor network \cite{Montangero2018}. This allows for the application of standard TN methods for the calculation of the evolution of $\rho(t)$, see e.g.~Ref.~\cite{Schollwock2011, Paeckel2019}.

For all values of $p_{1}$ and $p_{2}$, starting from a fully-occupied initial state, $\rho(0) = n \otimes n \otimes ... \otimes n$, the evolution of $\rho(t)$ up to time $T=100$ was estimated using MPS with fixed bond-dimensions of $\chi = 256, 512$ and for system sizes $L=64, 124$. For the special case of $p_{1} = 1$, an additional system size of $L=256$ was also used for the same bond-dimensions. From the simulation of the state $\rho(t)$, we compute the time evolution of the density  $\langle n(t)\rangle$ for each given pair of $p_{1}, p_{2}$. The values of $\langle n(t)\rangle$ at the latest time, $T=100$, are plotted on Fig.~3 of the main text. For the parameter grid, $10$ values of $p_{1}$ were chosen: $p_{1} \in \lbrace 0.1, 0.2, 0.3, 0.4, 0.5, 0.6, 0.7, 0.8, 0.9, 1.0 \rbrace$. For $p_{2}$ a finer grid of $28$ points was chosen: $p_{2} \in \lbrace 0, 0.01, 0.02, 0.03, 0.035, 0.04, 0.045, 0.05, 0.055, 0.06, 0.065, 0.07, 0.075, 0.08, 0.085, 0.09, 0.095, 0.1$, $0.1125, 0.125, 0.15, 0.1625, 0.175, 0.1875, 0.2, 0.2125, 0.225, 0.25\rbrace$. When combined, this produced a parameter grid of $280$ different points. For the special case of $p_{1} = 1$ with $L=256$, the chosen values of $p_{2}$ were: $p_{2} \in \lbrace 0.1625,0.175,0.1875,0.2,0.2125,0.225,0.25 \rbrace$.

\subsection{Estimate of the critical point}

\begin{figure}[t]
\centering
\includegraphics[width=0.7\linewidth]{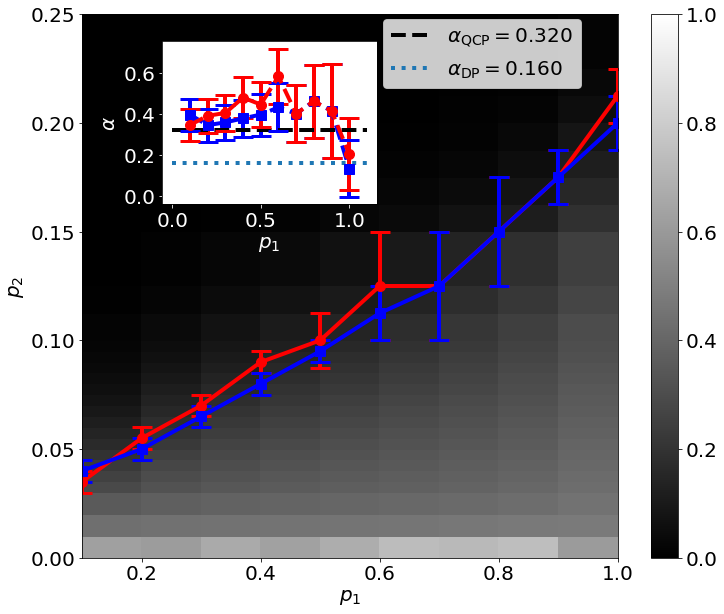}
\caption{\textbf{Tensor-network phase diagram:} The steady state density  is estimated by evolving $\rho(t)$ over $T=100$ steps using the method from Ref.~\cite{Gillman2021b}, with a bond-dimension of $\chi=512$, lattice size of $L=128$ and starting from a fully occupied initial state. The estimated critical line shown in red is estimated by taking, for each value of $p_{1} = 0.1, 0.2, ..., 1$, linear fits of $\langle n(t)\rangle$ for each value of $p_{2}$ over $t \in [10, 100]$. The best linear fit, as measured by the largest $R^{2}$ value, is then taken as the estimated critical point. The estimated critical line shown in blue instead uses the effective exponent, $\alpha(t)$. In this case, the location of the line closest to a constant provides the estimate of the critical point. This is measured by taking the minimum absolute gradient of $\alpha(t)$ averaged over $t \in [80,100]$. See the text for details. The inset shows the critical exponents, obtained by averaging $\alpha(t)$ over $t \in [80, 100]$, with the effective exponent taken at the correspond (red/blue) estimates of the critical point. Errors on each effective exponent separately are indicated by the error bars, and involve contributions from finite-sizes, finite bond-dimensions and finite parameter grid resolutions, see the text for details.}
\label{fig_sm:tn_pd}
\end{figure}

At the critical point, curves of $\langle n(t)\rangle$ obey power-law decays at sufficiently large $t$, i.e., $\langle n(t)\rangle \sim t^{-\alpha}$. As such, on a log-log plot these curves look  linear for large $t$. A linear-fit can thus be used to estimate the location of the critical point by looking at which curve is better approximated by a straight line in the log-log plot. For each value of $p_{1}$ all values of $p_{2}$ were considered and a linear fit produced over $t \in [10, 100]$. The best-linear-fit, as measured by the $R^{2}$ value, was selected as the estimated critical point. This produces the (estimated) critical line shown in Fig.~\ref{fig_sm:tn_pd} (red circles). Errors on this estimate are given by the grid resolution, where we have taken the maximum resolution from above and below to provide a single value for the error estimate.

Another way to estimate the critical line is by exploiting the  effective exponent $\alpha(t)$, defined as $\alpha(t) = - \log_{2}\left[\langle n(2 t)\rangle /\langle n(t)\rangle \right]$. For critical curves (i.e.~for real power-law behavior in $t$) this effective exponent is constant. The location of the critical line can then be estimated by selecting those parameters for which the effective exponent is the closest to a constant. For each value of $p_{1}$, the effective exponent was calculated for all corresponding values of $p_{2}$. The critical point was then estimated by taking the effective exponent with the minimum absolute gradient, averaged over $t \in [80,100]$. To ensure a fair comparison, before calculating the gradient each effective exponent is shifted by subtracting their mean value over $t \in [80, 100]$, i.e., by the estimated value of the critical exponent itself. This ensures all curves average around zero in the examined window. Additionally, for $p_{1} = 1$, only $p_{2} > 0.15$ were considered. This is necessary since for low values of $p_{2}$ the phase diagram is deep into the active phase, and this causes the effective exponents to rapidly vanish. The final estimated critical curve resulting from this procedure is shown in Fig.~\ref{fig_sm:tn_pd} (blue squares). The errors on these estimates are again given by the finite resolution of the grid, taking the maximum absolute difference of the next values of $p_{2}$ above and below the estimated critical value to provide the error estimate.

Averaging the two critical lines obtained through the two methods discussed above leads to the estimated critical line shown in Fig.~3 of the main text, with the error bars combining the two separate error estimates via a root sum-of-squares.

\subsection{Estimate of the critical exponent}

\begin{figure}[t]
\centering
\includegraphics[width=1\linewidth]{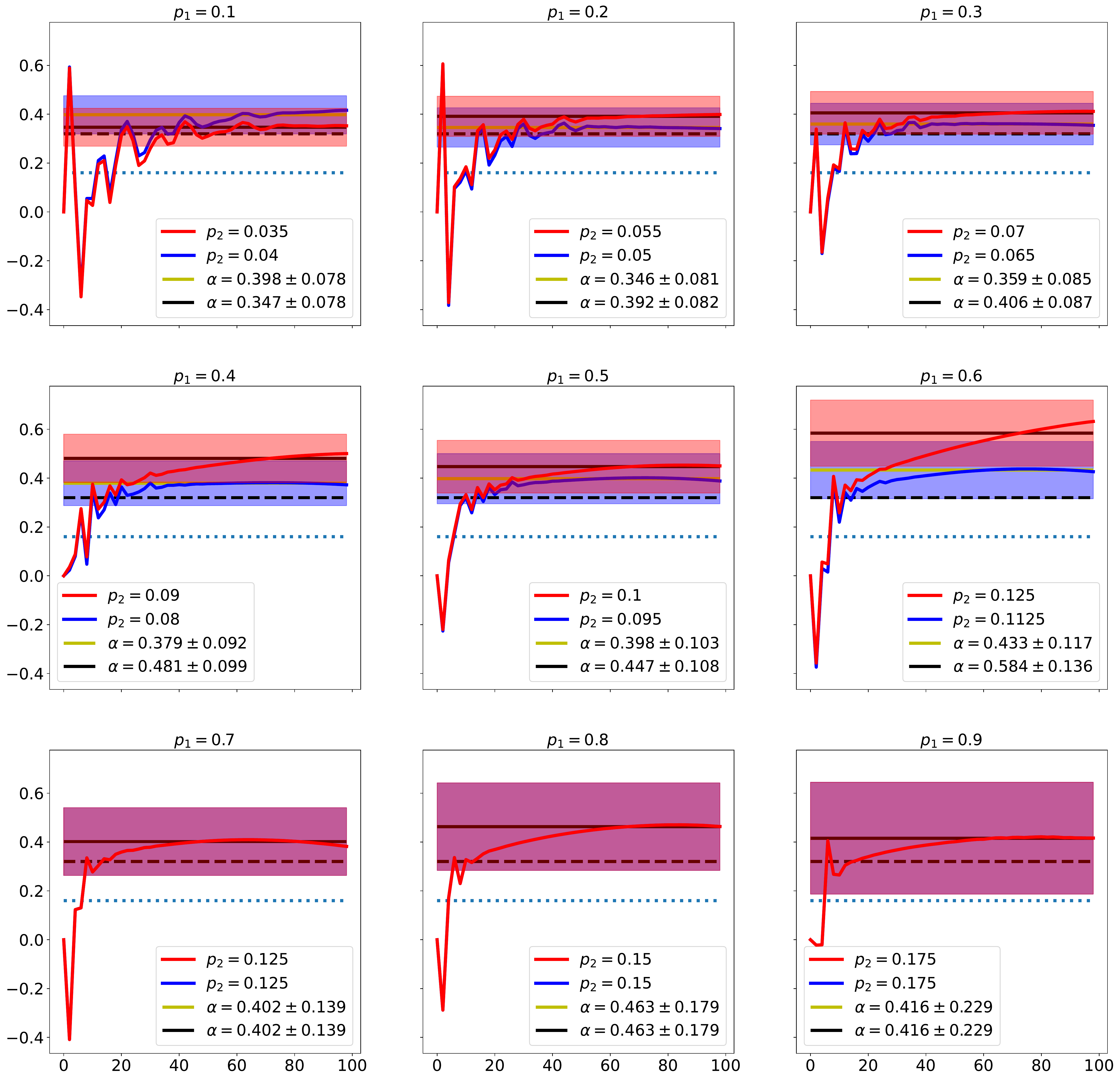}
\caption{\textbf{Effective exponent:} The effective exponent $\alpha(t)$ for the QCA with $p_{1} \in [0.1, 0.9]$, along with the values of $\alpha$ for the $1D$ QCP (dashed black line, $\alpha = 0.32$) and for $1D$ DP (dotted line, $\alpha=0.16$). Here, $\alpha(t)$ is shown for both estimated critical points in Fig.~\ref{fig:tn_pd} (red for the linear-fit method and blue for the effective exponent method). The two estimates for the critical exponents, obtained by averaging $\alpha(t)$ over $t \in [80,100]$, are indicated by solid horizontal lines. Errors occurring on each critical exponent estimate (corresponding to the error bars in the inset of Fig.~\ref{fig_sm:tn_pd}) are indicated by the shaded regions and involve contributions from finite-sizes, finite bond-dimensions and finite parameter grid resolutions, see text for details. These errors, particularly those associated to finite-size, tend to increase with $p_{1}$.}
\label{fig_sm:effective_exponents}
\end{figure}

The effective exponent $\alpha(t)$ used above for finding the critical line can also be exploited to estimate the critical exponent $\alpha$. For each $p_{1}$ and corresponding estimated critical value of $p_{2}$, the effective exponent was averaged over a window of $t \in [80,100]$ to estimate $\alpha$. The plots of the effective exponents themselves are shown in Fig.~\ref{fig_sm:effective_exponents} for $p_{1} \in [0.1, 0.9]$. Estimates for each critical point are indicated in the inset of Fig.~\ref{fig_sm:tn_pd}, as well as by horizontal lines in Fig.~\ref{fig_sm:effective_exponents}.

To calculate error estimates on the critical exponents, we have considered several error sources. First, finite-size effects were quantified by taking the absolute difference between $\alpha$ calculated with $L=124, \chi=512$ (recalling that $\alpha$ is calculated as an average of $\alpha(t)$ over $t \in [80, 100]$) and $\alpha$ calculated with $L=64, \chi=512$, i.e., with half the lattice size. Similarly, finite bond-dimension effects were quantified by the absolute difference between $\alpha$ calculated with $L=124, \chi=512$ and $\alpha$ calculated with $L=124, \chi=256$, i.e., with half the bond-dimension. To estimate the error due to the parameter grid resolution, for each value of $p_{1}$ the maximum absolute difference was taken between a given critical exponent estimate, $\alpha(p_{1}, p_{2})$, and the values of $\alpha$ calculated from the closest surrounding values of $p_{2}$. For example, for $p_{1} = 0.2$ and $p_{2} = 0.035$ one calculates $|\alpha(p_{1}=0.2, p_{2}=0.035) - \alpha(p_{1}=0.2, p_{2}=0.04)|$. This is repeated for the value of $p_{2}$ just below the one of interest, and the maximum of these two is taken as the error estimate. If the there is no value above (below) the chosen value of $p_{2}$, then only the value below (above) is used for the estimate. These three errors are then combined as fractional errors via a root sum-of-squares, to give a single error estimate for the critical exponent. 

\begin{figure}[t]
\centering
\includegraphics[width=0.7\linewidth]{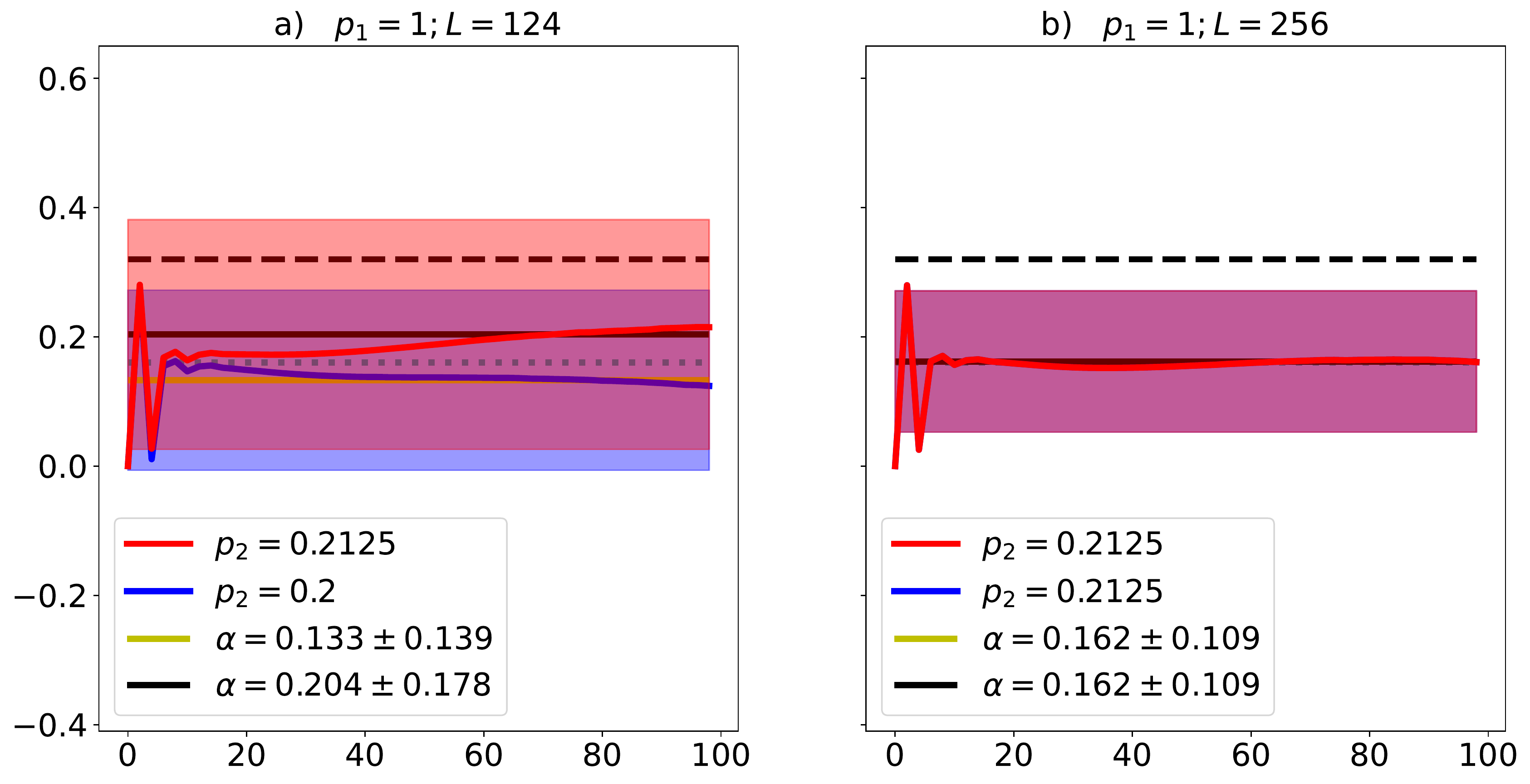}
\caption{\textbf{Effective exponents for $p_{1} = 1$:} 
The effective exponent for $p_{1} = 1$ with $L=124$ and $L=256$ for panels a) and b) respectively. See the caption of Fig. \ref{fig_sm:effective_exponents} for more details on the plots. As can be seen, increasing the system size dramatically reduces the errors, and the resulting estimate of $\alpha$ for $p_{1} = 1$ (solid black line) lies close to the $1D$ DP value of $0.16$ (dotted line)}
\label{fig_sm:effective_exponents_DP}
\end{figure}

The errors on the estimated values of $\alpha$ are displayed via error bars in the inset of Fig.~\ref{fig_sm:tn_pd}. They are also indicated by the shaded regions in Fig.~\ref{fig_sm:effective_exponents}. As can be seen, the size of errors tends to increase with $p_{1}$. This is largely attributable to finite-size effects, which can be expected to increase with the probability of sites being occupied. As such, while for low values of $p_{1}$, the estimated critical exponents lie, well within error bars, close to the value of the quantum contact process, for higher $p_{1}$ this can become ambiguous. To gain a more precise estimate of the value of $\alpha$ for the extreme case of $p_{1} = 1$, we then repeat the above procedure but with simulations of $L=256$ and a grid of $p_{2} \in \lbrace 0.1625,0.175,0.1875,0.2,0.2125,0.225,0.25 \rbrace$. This leads to the effective exponents shown in the rightmost panel of Fig.~\ref{fig_sm:effective_exponents_DP} (b), with the original estimate using $L=124$ show in the panel (a) for comparison.

Finally, to obtain a single estimate of the critical exponent $\alpha$, we take an average of the two estimates arising from each estimate of the critical point. The final error estimate is then produced by taking the root sum-of-squares of the two individual error estimates. The resulting critical exponents are collected in Table S1.  

\begin{table}
	\begin{center}
		\begin{tabular}{c|c|c|c}
			Model & $p_1$ & $p_2$ & $\alpha$ \\
			\hline
			CCP/DKCA (DP) \cite{Henkel2008} & - & - & $0.159464 \pm 0.000006$ \\
			QCP \cite{Jo2021} & - & - & $0.32\pm 0.01$ \\
			QCP \cite{Carollo2019} & - & - & $0.36 \pm 0.08$ \\
			QCA	 & 0.1 & 0.038 $\pm$ 0.007 & 0.37 $\pm$ 0.11 \\
			 & 0.2 & 0.053 $\pm$ 0.007 & 0.37 $\pm$ 0.12 \\
			 & 0.3 & 0.068 $\pm$ 0.007 & 0.38 $\pm$ 0.12 \\
			 & 0.4 & 0.085 $\pm$ 0.007 & 0.43 $\pm$ 0.14 \\
			 & 0.5 & 0.098 $\pm$ 0.013 & 0.42229 $\pm$ 0.14897 \\
			 & 0.6 & 0.119 $\pm$ 0.028 & 0.509 $\pm$ 0.179 \\
			 & 0.7 & 0.125 $\pm$ 0.035 & 0.402 $\pm$ 0.196 \\
			 & 0.8 & 0.15 $\pm$ 0.04 & 0.46 $\pm$ 0.25 \\
			 & 0.9 & 0.175 $\pm$ 0.018 & 0.416 $\pm$ 0.324 \\
			 & 1 & 0.206 $\pm$ 0.018 & 0.17 $\pm$ 0.23 \\
			QCA (L=256) & 1 & $0.2125 \pm 0.0125$ & $0.16 \pm 0.11$ \\
		\end{tabular}
	\end{center}
	\caption{{\bf Table of critical exponents}. The value of $\alpha$ for one-dimensional directed percolation (DP), as e.g. encapsulated in models of the classical contact process (CCP) or Domany-Kinzel cellular automata (DKCA), is shown in the first row. The second and third row show estimates of $\alpha$ for the continuous-time quantum contact process (QCP). These can be compared with the esimates of $\alpha$ obtained for the QCA in the main text, across the phase diagram. For low $p_{1}$ these tend to agree with $\alpha$ for the QCP. For $p_{1} = 1$, an estimate of $\alpha$ with a larger lattice size of $L=256$ is consistent with the $1D$ DP value. In the intermediate range of $p_{1}$ values, errors become sufficiently large with $p_{1}$ so as to make conclusions difficult. This is compounded by the suggestion in Ref.~\cite{Jo2021} that, in fact, the QCP can experience a continuous change of critical exponent when classical effects are included moving from the QCP value of $\alpha$ to that of $1D$ DP when the classical effects are sufficiently large.}
\end{table}

\end{document}